\documentclass[acmtog,nonacm]{acmart}   
\author{Manas Chaudhary}
\authornote{Both authors contributed equally.}
\affiliation{%
  \institution{Indian Institute of Technology Delhi} \country{India}}
\email{manaschaudhary2000@gmail.com}

\author{Chandradeep Pokhariya}
\authornotemark[1] 
\affiliation{%
  \institution{Indian Institute of Technology Delhi} \country{India}}
\email{cdpokhariya@gmail.com}

\author{Rahul Narain}
\affiliation{%
  \institution{Indian Institute of Technology Delhi} \country{India}}
\email{narain@cse.iitd.ac.in}

\usepackage{float}

\usepackage{booktabs} 
\usepackage[ruled]{algorithm2e} 

\SetAlFnt{\small}
\SetAlCapFnt{\small}
\SetAlCapNameFnt{\small}
\SetAlCapHSkip{0pt}

\newcommand{\bfd}{{\mathbf d}}

\newcommand{\bff}{{\mathbf f}}

\newcommand{\bfs}{{\mathbf s}}

\newcommand{\bfv}{{\mathbf v}}

\newcommand{\bfx}{{\mathbf x}}

\newcommand{\bfE}{{\mathbf E}}
\newcommand{\bfF}{{\mathbf F}}

\newcommand{\bfH}{{\mathbf H}}
\newcommand{\bfI}{{\mathbf I}}

\newcommand{\bfM}{{\mathbf M}}

\newcommand{\bfS}{{\mathbf S}}

\newcommand{\bfW}{{\mathbf W}}

\newcommand{\bfzero}{{\mathbf 0}}

\usepackage{bm}
\usepackage{upgreek}

\newcommand{\bflambda}{{\bm\uplambda}}

\newcommand{\rmd}{{\mathrm d}}

\newcommand{\rms}{{\mathrm s}}

\newcommand{\bbR}{{\mathbb R}}

\newcommand{\Dt}{\Delta t}

\begin{document}

\title{Towards Generalized Position-Based Dynamics}

\newcommand{\TODO}[1]{\textbf{\color{red}[TODO: #1]}}

\begin{abstract}
  The position-based dynamics (PBD) algorithm is a popular and versatile technique for real-time simulation of deformable bodies, but is only applicable to forces that can be expressed as linearly compliant constraints.
  In this work, we explore a generalization of PBD that is applicable to arbitrary nonlinear force models.
  We do this by reformulating the implicit time integration equations in terms of the individual forces in the system, to which applying Gauss-Seidel iterations naturally leads to a PBD-type algorithm.
  As we demonstrate, our method allows simulation of data-driven cloth models \cite{sperl2020hylc} that cannot be represented by existing variations of position-based dynamics, enabling performance improvements over the baseline Newton-based solver for high mesh resolutions.
  We also show our method's applicability to volumetric neo-Hookean elasticity with an inversion barrier.
\end{abstract}



\begin{teaserfigure}
    \centering
    \includegraphics[width=\textwidth]{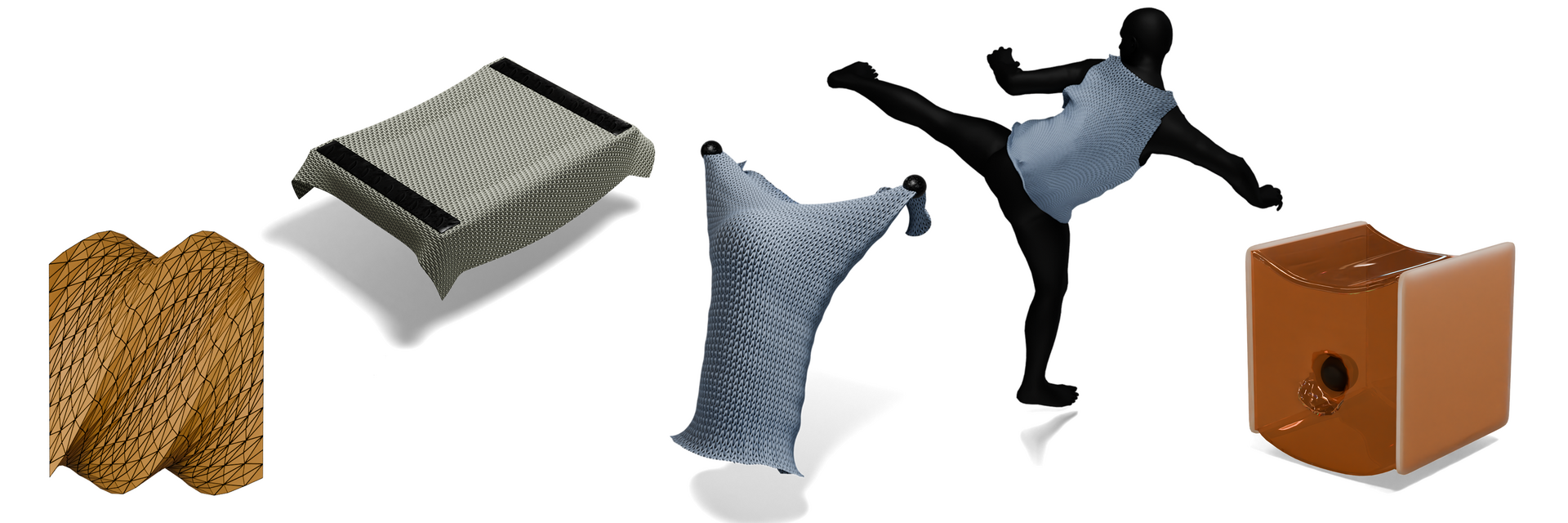}
    \caption{Our generalization of position-based dynamics allows simulation of data-driven cloth model from \cite{sperl2020hylc} and neo-Hookean with an inversion barrier while demonstrating the expected material behavior.}
    \label{fig:teaser}
\end{teaserfigure}%


\maketitle

\section{Introduction}
\label{sec:introduction}

Real-time simulation of highly deformable materials such as cloth has long been one of the major goals of computer graphics.
It remains a challenging problem due to the need for robust, stable time integration in the presence of unpredictable user input and a tightly constrained computational budget, especially if one wishes to depict realistic materials which have a highly nonlinear elastic response.
In offline simulation, robustness and stability is ensured by implicit time integration schemes such as backward Euler.
However, these methods require the solution of a large system of nonlinear equations at each time step, often necessitating multiple iterations of Newton's method to arrive at a converged solution.

Since such a strategy is far too expensive for interactive applications, many techniques have been proposed in the graphics literature to reduce the computational cost \cite{bender2017survey,bouaziz2014projective,liu2017quasinewton,lan2023stencil,chen2024vbd}.
In this work, we focus on position-based dynamics (PBD) \cite{bender2017survey}, which is extremely popular in interactive applications due to its stability, performance, and ease of implementation.
Unlike most other iterative techniques used for dynamics simulation, PBD is a so-called ``dual'' descent method \cite{macklin2020pd}: it iterates over the forces in the system, and updates the effect of each force one at a time, rather than updating vertex positions as in ``primal'' methods.
As shown by \citet{macklin2020pd}, this has the advantage that PBD is better able to handle large stiffness ratios between competing forces in the system compared to primal methods, although it is worse at handling large mass ratios.

A fundamental limitation of PBD-based methods is that all forces must be modeled as linearly compliant constraints \cite{macklin2016xpbd}.
That is, the force must be associated with a fixed constraint function, and its magnitude must be linearly proportional to the constraint violation.
This precludes the simulation of many realistic material models, whose elastic response is often highly nonlinear.
Our goal in this work is to remove this limitation, and allow arbitrary energy-based models to be expressed in the position-based dynamics framework.
To enable PBD simulation of data-driven cloth models, we reformulate PBD in a way that supports arbitrary energy-based models.
To do so, we express implicit time integration as a system of equations whose unknowns are the displacements caused by all the forces acting on the system.
Applying Gauss-Seidel or Jacobi iterations on this formulation leads to an algorithm equivalent to extended position-based dynamics \cite{macklin2016xpbd} for the linearly compliant case.
In general, we iterate over each individual force term in the system (for example, the force due to one element on its adjacent vertices in a FEM mesh), and update it implicitly using a low-dimensional Newton solve on the element strain.

Our proposed scheme is expressive enough to apply to any nonlinear energy-based elastic model, and we demonstrate its application to two realistic material models that have so far been out of reach of PBD simulation.
First, we simulate of neo-Hookean elastic bodies with a $\log|\bfF|$ inversion barrier, whose strain energy rises to infinity when compressed to zero volume.
Second, we simulate cloth with a homogenized yarn-level model \cite{sperl2020hylc}, which relies on a precomputed database capturing the nonlinear stretching and bending behaviour of fabric.
Our algorithm retains the flexibility of PBD, for example easily incorporating various other constraints and collisions as additional projection steps in the solver loop.
Our GPU implementation is parallelized using graph colouring and scales well with increasing mesh resolution.



\section{Related work}
\label{sec:related}

There has been a great deal of work on algorithms for real-time simulation for deformable bodies in the graphics literature.
Techniques based on position-based dynamics \cite{bender2017survey} model all forces as linearly compliant constraints, and compute the corresponding Lagrange multipliers in an iterative fashion.
Projective dynamics \cite{bouaziz2014projective,liu2017quasinewton} makes a similar assumption, but performs alternating optimizations on position coordinates and constraint projections.
Many other descent methods such as stencil descent \cite{lan2023stencil} and vertex block descent \cite{chen2024vbd} have also been proposed recently.
For brevity, we focus on methods based on position-based dynamics, which are the most closely related to our work.


Position-based dynamics (PBD) \cite{mueller2006pbd} was introduced as a dynamics formulation in which all interaction forces are interpreted as constraints.
The algorithm proceeds by iteratively satisfying constraints one at a time, by updating vertex positions in a Gauss-Seidel or Jacobi manner.
Since its introduction, PBD has been applied to a wide variety of phenomena, including deformable bodies \cite{diziol2011incompressible,bender2014continuous,mueller2014strain,macklin2021neohookean,chen2023hyperelasticity}, fluids \cite{macklin2013fluids,xing2022surface}, rigid and articulated bodies \cite{deul2014rigid,mueller2020rigid}, elastic rods \cite{umetani2014rods,kugelstadt2016rods}, and crowds \cite{weiss2019crowds}.
A unified framework based on PBD \cite{macklin2014unified} allows two-way interaction between diverse phenomena such as fluids, rigid bodies, cloth, etc.\ in real time.
Most recently, \citet{yu2024xpbi} combine PBD with an implicit plasticity update to simulate elastoplastic behaviour.
A survey of many such techniques is given by Bender et al.~\cite{bender2017survey}.

The original PBD formulation \cite{mueller2006pbd} only supported hard constraints, with apparent compliance and elasticity arising because the iterations were terminated before convergence.
As a result, the effective stiffness of the simulated materials would depend on the iteration count.
The extended position-based dynamics (XPBD) technique of \citet{macklin2016xpbd} removed this limitation, providing a consistent formulation for compliant (soft) constraints.
More recent work has advocated for the use of substepping and lower iteration counts over large time steps and more iterations \cite{macklin2019small}.

To model continuum elasticity in the PBD framework, a number of approaches have been proposed.
Early techniques \cite{bender2014continuous,mueller2014strain} suffered from the iteration-dependent stiffness.
Only recently have methods appear that are capable of faithfully modeling particular hyperelastic material models.
In particular, the stable neo-Hookean model \cite{smith2018flesh} has been shown to be expressible in a PBD-compatible form \cite{macklin2021neohookean}.
Closest to our work, Chen et al.~\cite{chen2023hyperelasticity} simulate arbitrary hyperelastic materials by solving for the first Piola-Kirchhoff stress, rather than the constraint Lagrange multiplier, at each element.

Other than formulations for modeling particular phenomena in the PBD framework, we also note model-independent techniques for speeding up the PBD solves themselves.
Graph colouring \cite{fratarcangeli2016vivace} and clustering \cite{tonthat2023parallel} can be used to improve the parallelism of Gauss-Seidel updates.
The convergence rate of the iterations can be increased by acceleration techniques such as the Chebyshev semi-iterative method \cite{wang2015chebyshev} and Anderson acceleration \cite{peng2018anderson}.
Such techniques could be applicable to our work as well.

Despite the versatility of the PBD framework, almost all prior instantiations of this approach require the simplifying approximation that all forces must be expressed as constraints and computed through Lagrange multipliers.
As a result, PBD simulations are unable to model arbitrary interaction models, such as data-driven stress-strain relationships.
To our knowledge, the FP-PXPBD method of \citet{chen2023hyperelasticity} is the only work that lifts this restriction, but it focuses only on finite element elasticity.
On the other hand, we propose a general framework that applies to generic discretizations including cloth simulation.

\section{Generalizing position-based dynamics}
\label{sec:theory}

We consider a physical system spatially discretized into $n$ particles having positions $\bfx = (\bfx_1,\dots,\bfx_n) \in \bbR^{3n}$ and velocities $\bfv = (\bfv_1,\dots,\bfv_n) \in \bbR^{3n}$.
As usual, we define the mass matrix $\bfM = \operatorname{diag}(m_1\bfI,\dots,m_n\bfI) \in \bbR^{3n\times3n}$.
The elastic behaviour of the body is determined by its elastic energy, given as the sum of several energy terms $U_i(\bfx)$, resulting in forces $\bff_i(\bfx)=-\nabla U_i(\bfx)$.
In addition there may be numerous other forces acting on the particles, for example gravity, collisions, constraints, wind, and so on, all of which may be modeled as additional energy terms or as external forces whose sum is $\bff_{\text{ext}}$.

Let us denote the positions and velocities at the beginning of the time step as $\bfx^-$ and $\bfv^-$, while those at the end are simply $\bfx$ and $\bfv$.
Performing implicit time integration can be viewed as finding the new state $(\bfx,\bfv)$ which is consistent with the forces evaluated there. For example, the backward Euler scheme is
\begin{align}
  \bfx &= \bfx^- + \Dt\bfv^- + \Dt^2\bfM^{-1}\bigl(\bff_{\text{ext}}+\sum_i\bff_i(\bfx)\bigr), \\
  \bfv &= \tfrac1{\Dt}(\bfx - \bfx^-).
\end{align}
Many other implicit schemes such as the trapezoidal rule, BDF2, and so on can be expressed in a similar way.
Since $\bfv$ can be easily recovered from $\bfx$, from here on we focus on computing $\bfx$.

We can simplify the discretized equations of motion by defining
\begin{align}
  \tilde\bfx &:= \bfx^- + \Dt\bfv^- + \Dt^2\bfM^{-1}\bff_{\mathrm{ext}}, \\
  \bfW &:= \Dt^2\bfM^{-1}, \\
  \bfd_i &:= \bfW\bff_i(\bfx),
\end{align}
where $\tilde\bfx$ is the predicted position in the absence of internal forces, $\bfW$ is proportional to the inverse mass, and $\bfd_i$ is the displacement caused by each force.
Thus, we obtain a new system of equations
\begin{align}
  \bfx &= \tilde\bfx + \sum_i\bfd_i, \\
  \bfd_i &= -\bfW\nabla U_i(\bfx) \quad \text{for } i=1,2,\dots
\end{align}
As we will see, position-based dynamics can be viewed as a way to iteratively update $\bfd_i$ for each force, and thus obtain the solution $\bfx$.

In practice, the internal energy terms $U_i$ usually depend only on lower-dimensional quantity $\bfs_i(\bfx)$.
For example, the energy of a spring depends only on the distance between its two endpoints, the bending energy of a hinge element depends only on the angle between its adjacent faces, and the stretching energy of a triangle depends only on the three independent entries of its Green strain tensor.
Thus we have a reduced energy $U_i(\bfx)=\hat U_i(\bfs_i(\bfx))$ and the force is of the form $\bff_i = -\bfS_i(\bfx)^T\nabla\hat U_i$ where $\bfS_i=\rmd\bfs_i/\rmd\bfx$ is the Jacobian of the strain.

\subsection{XPBD}
\label{sec:xpbd}

In this section, we review the popular XPBD algorithm \cite{macklin2016xpbd}.
XPBD only permits energies that are expressed as
\begin{align}
  U_i(\bfx) = \frac1{2\alpha}c_i(\bfx)^2,
  \label{eq:xpbd-energy}
\end{align}
where $\alpha\ge0$ is the compliance and the constraint function $c_i(\bfx)$ is scalar-valued.
Comparing with our formulation, $c_i$ plays the role of the strain $\bfs_i$, and the reduced energy is simply $\hat U(s) = \frac1{2\alpha}s^2$, giving $\bff_i=-\frac1\alpha c_i\nabla c_i$.

The XPBD algorithm maintains a Lagrange multiplier $\lambda_i$ corresponding to each constraint, so that the force due to the constraint is taken to be $\bff_i = \Dt^{-2}\lambda_i\nabla c_i(\bfx)$.
It proceeds by iterating over each energy term $i$, and updating the particle positions $\bfx$ to satisfy the constraint:
\begin{align}
  \lambda_i &\gets \lambda_i + \Delta\lambda, \label{eq:xpbd-lambda-update} \\
  \bfx &\gets \bfx + \bfM^{-1}\Delta\lambda\nabla c_i(\bfx), \label{eq:xpbd-x-update}
\end{align}
where
\begin{align}
  \Delta\lambda&=-\frac{c_i(\bfx)+\Dt^{-2}\alpha_i\lambda_i}{\nabla c_i^T\bfM^{-1}\nabla c_i+\Dt^{-2}\alpha_i} \label{eq:xpbd-delta-lambda}
\end{align}
is the change in the Lagrange multiplier such that the updated positions satisfy the constraint equations.
This update rule is derived under the assumption that the constraint gradients $\nabla c_i(\bfx)$ are not changing significantly across iterations.

There are two commonly used strategies for applying these per-element updates on a simulation mesh.
In a parallel Gauss-Seidel approach, graph colouring is performed on the dual graph consisting of elements sharing common vertices.
All elements with the same colour can then be updated in parallel.
In a Jacobi approach, the updates are computed in parallel across all elements of the mesh, and the displacements due to each element update are averaged to increment the vertex positions.
Since averaging slows down the convergence of the method, the displacements are usually multiplied by an over-relaxation factor $\omega \in [1, 2)$ when updating positions.

To make the connection with our method, let us explicitly keep track of the displacements $\bfd_i$ caused by each force.
In particular, we can update
\begin{align}
  \bfd_i &\gets \bfd_i + \bfM^{-1}\Delta\lambda\nabla c_i(\bfx),
\end{align}
and simply have $\bfx = \tilde\bfx + \sum_i\bfd_i$.
If the assumption that $\nabla c_i$ is a constant throughout the iterations holds, \eqref{eq:xpbd-lambda-update} implies that $\bfd_i = \bfM^{-1}\lambda_i\nabla c_i$ at any point in the algorithm.
In that case, we show in Appendix \ref{app:equivalence} that performing a Newton update on $\bfd_i$ to satisfy its force equation:
\begin{align}
  \bfd_i &\gets \bfd_i + \Delta\bfd_i, \label{eq:gpbd-d-update} \\
  \bfx &\gets \bfx + \Delta\bfd_i, \label{eq:gpbd-x-update} \\
  \text{s.t. }\bfd_i+\Delta\bfd_i &= -\Dt^2\bfM^{-1}\nabla U_i(\bfx+\Delta\bfd_i) \label{eq:gpbd-eq}
\end{align}
is equivalent to the XPBD update rule, i.e.\ it leads to the update $\Delta\bfd_i = \bfM^{-1}\Delta\lambda\nabla c_i$ where $\Delta\lambda$ is given by \eqref{eq:xpbd-delta-lambda}.

\subsection{Generalized PBD}
\label{sec:gpbd}

We now introduce our generalization of the XPBD approach to energy terms that are not of the form \eqref{eq:xpbd-energy}.
As before, for each energy term $i$, we assume that all other forces and their resulting displacements $\bfd_j$ are fixed, and update only $\bfd_i$.
We will also maintain the invariant that our current guesses for $\bfx$ and $\bfd_i$'s always satisfy $\bfx = \tilde\bfx + \sum_i\bfd_i$.
So, we simply have to update $\bfd_i$ according to \eqref{eq:gpbd-d-update}--\eqref{eq:gpbd-eq}, which is a well-defined problem for arbitrary energy functions $U_i$.

The main computational task to solve \eqref{eq:gpbd-eq}.
To this end, it is helpful to observe that this equation is the optimality condition for a variational problem,
\begin{align}
  \min_{\Delta\bfd_i} \frac12\|\bfd_i+\Delta\bfd_i\|_{\bfW^{-1}}^2 + U_i(\bfx+\Delta\bfd_i),
  \label{eq:gbpd-opt}
\end{align}
where $\|\cdot\|_{\bfW^{-1}}$ denotes the weighted norm $\bfd \mapsto \sqrt{\bfd^T\bfW^{-1}\bfd}$.




While in principle $\bfd_i$ is $3n$-dimensional like $\bfx$, it is not necessary to carry out this optimization over all $3n$ degrees of freedom.
If $U_i$ is only a function of a $k$-dimensional strain measure $\bfs_i\in\bbR^k$, then the force $\bff_i$ must lie in a $k$-dimensional subspace, namely the row space of $\bfS_i=\rmd\bfs_i/\rmd\bfx$:
\begin{align}
  \bff_i = -\nabla U(\bfx) = -\bfS_i(\bfx)^T\nabla\hat U_i.
\end{align}
Similarly, the displacement $\bfd_i=\bfW\bff_i$ should lie in the column space of $\bfW\bfS_i(\bfx)^T$.
In the spirit of XPBD, we only restrict the incremental displacement $\Delta\bfd_i$ to lie in this $k$-dimensional subspace, by taking it to be of the form
\begin{align}
  \Delta\bfd_i = \bfW\bfS_i(\bfx)^T\Delta\bflambda
\end{align}
for some unknown vector $\Delta\bflambda\in\bbR^k$.
Then we perform minimization over this lower-dimensional subspace,
\begin{align}
  \min_{\Delta\bflambda}\frac12\|\bfd_i+\bfW\bfS_i^T\Delta\bflambda\|_{\bfW^{-1}}^2 + U_i(\bfx+\bfW\bfS_i^T\Delta\bflambda), \label{eq:gpbd-reduced-opt}
\end{align}
with $\bfS_i$ evaluated at the current iterate $\bfx$ and held fixed.
We do not attempt to enforce that the full updated displacement $\bfd_i+\Delta\bfd_i$ itself lie in the correct subspace.
In our experiments, we found that this improved stability, albeit at the cost of preventing convergence to the exact backward Euler solution.

This approach preserves and generalizes the low-dimensional projections used in traditional XPBD.
For example, if the energy is of the form \eqref{eq:xpbd-energy}, we can define the 1-dimensional strain $\bfs_i = c_i$, so that $\bfS_i^T=\nabla c_i$ and \eqref{eq:gpbd-reduced-opt} becomes a 1D problem with a solution similar to \eqref{eq:xpbd-delta-lambda}.
For a triangular finite element whose stretching energy depends on its Green strain tensor $\bfE=\frac12(\bfF^T\bfF-\bfI)$, we can take its three independent entries $\bfs_i=(e_{xx},e_{xy},e_{yy})$, and thus we only have to perform a 3-dimensional optimization instead of optimizing all 9 DOFs of the triangle vertices.



The resulting algorithm is summarized in Algorithm~\ref{alg:gpbd}.
We emphasize that our scheme is easy to integrate into an existing (X)PBD implementation, since the overall structure of the PBD loop is unchanged: in each iteration, we iterate over each energy (equivalently, each constraint, force, etc.) in the system, and perform a projection to update the current configuration $\bfx$ according to it.
For hard constraints and quadratic energies, we can continue to perform PBD or XPBD-style projections, while general nonlinear energies can be updated using our approach.
For example, we can perform two-way collision handling using PBD projections, as shown in Section~\ref{sec:results}.

\begin{algorithm}[t]
\SetAlgoLined
\KwData{initial positions $\bfx^-$ and velocities $\bfv^-$, timestep size $\Dt$}
\KwResult{final positions $\bfx$ and velocities $\bfv$}
initialize $\bfx = \bfx^- + \Dt\bfv^- + \bfW\bff_{\text{ext}}$\;
initialize $\bfd_i = \bfzero$ for all $i=1,2,\dots$\;
\For{GPBD iterations $k=1,2,\dots$}{
  \For{each force $i=1,2,\dots$} {
    evaluate $\bfS_i = \rmd\bfs_i/\rmd\bfx$ at the current state $\bfx$\;
    compute $\Delta\bflambda$ to minimize \eqref{eq:gpbd-reduced-opt}\;
    compute $\Delta\bfd_i = \bfW\bfS_i^T\Delta\bflambda$\;
    increment $\bfd_i$ and $\bfx$ by $\Delta\bfd_i$\;
  }
}
set $\bfv = \Dt^-(\bfx - \bfx^-)$\;
\caption{The GPBD algorithm. For ease of exposition, we present the algorithm in Gauss-Seidel form; in practice, we use Jacobi with over-relaxation for improved parallelizability.}
\label{alg:gpbd}
\end{algorithm}

\section{Applications}

\subsection{Hyperelastic finite elements}
\label{sec:fem}
Our method supports finite element simulation of arbitrary hyperelastic materials.
Here we focus on two cases: neo-Hookean elasticity with a hard inversion barrier \cite{sifakis2012fem}, and the stable neo-Hookean model \cite{smith2018flesh}.
In these models, the elastic potential of an tetrahedron with volume $V_j$ and deformation gradient $\bfF_j$ is given by
\begin{equation}
  U_j = V_j\left(\frac\mu2(\|\bfF_j\|^2-3) - \mu\log|\bfF_j| + \frac\lambda2\log^2|\bfF_j|\right) \label{eq:log-nh}
\end{equation}
and
\begin{equation}
  U_j = V_j\left(\frac\mu2(\|\bfF_j\|^2-3) - \mu(\bfF_j - 1) + \frac\lambda2(F_j - 1)^2\right)
\end{equation}
respectively.
In particular, the neo-Hookean model \eqref{eq:log-nh} is undefined under inversion, when $|\bfF_j|<0$.

In this case, we must take care not to initialize the per-element solver in such an inverted state.
We can treat such an element conceptually as a combination of a non-inversion constraint $|\bfF_j|>0$, and an elastic energy defined above.
The former is a hard nonlinear constraint and can be handled by vertex projection in the classic PBD manner, while the latter is handled by our GPBD update.
In particular, if the element is inverted, we find the smallest singular value of $\bfF$ and negate it, then update the vertex positions to achieve the new $\bfF$ while preserving the center of mass.
Then, we perform our GPBD update from the new vertex positions, using the symmetric entries of the Green strain $\bfE = \frac12(\bfF^T\bfF - \bfI)$ to perform a 6-dimensional minimization using Newton's method with line search.

This approach is simple to implement, and consistent with the overall PBD framework, since handling hard constraints with PBD does not require tracking any additional forces or Lagrange multipliers.
We note that such a strategy for handling inverted elements is only possible in a position-based setting, whereas descent methods would not be able to recover from the presence of a single inverted element.


\subsection{Data-driven cloth simulation}
\label{sec:hylc}

In this section, we discuss how to apply our generalized position-based dynamics algorithm to cloth simulation using the homogenized yarn-level cloth (HYLC) model of \citet{sperl2020hylc}.
This is a data-driven technique in which the elastic energy of a yarn-level cloth model is precomputed at various stretching and bending strains, and interpolated to the current configuration at runtime.
Specifically, the energy is a function of the first and second fundamental forms ($\bfI$ and ${\bf II}$) of the cloth mid-surface, discretized on triangular faces using adjacent face normals \cite{10.1145/3197517.3201395}.
This results in one internal energy term $U_i$ for each mesh triangle, combining both stretching and bending responses.

Both $\bfI$ and ${\bf II}$ are symmetric $2\times2$ matrices, depending on the positions of the three vertices of the triangle and the three additional vertices from the adjacent faces.
From these fundamental forms we form a 6-dimensional strain vector $\bfs_i$ by stacking the in-plane strain measures
\begin{align}
  s_x&=\sqrt{I_{xx}}-1, & s_a&=I_{xy}/\sqrt{I_{xx}I_{yy}}, & s_y&=\sqrt{I_{yy}}-1
\end{align}
from \citet{sperl2020hylc}, and the 3 independent entries $II_{xx},II_{xy},II_{yy}$ of ${\bf II}$.
All relevant energies, forces, and Jacobians are computed in terms of these strain measures using the original HYLC code \cite{sperl2020hylc}.
With this formulation, the GPBD update \eqref{eq:gpbd-reduced-opt} involves a 6-dimensional optimization.



Since the HYLC energies are highly nonlinear (as also observed by \citet{feng2024neural}), we make two additional considerations to help the solver make rapid progress in a small number of iterations.
First, we found that the term $U_i(\bfx+\bfW\bfS_i^T\Delta\bflambda)$ in \eqref{eq:gpbd-reduced-opt} becomes very ill-conditioned at high mesh resolutions, due to the different scaling properties of the stretching and bending modes.
This issue can be significantly improved simply by rescaling the components of $\bfs_i$.
In particular, in-plane strain quantities are dimensionless, while bending strains are curvatures and have units of 1/length.
This affects conditioning of the Hessian $\bfS_i(\frac{\rmd^2 U_i}{\rmd\bfx^2})\bfS_i^T$ in two ways: through the conditioning of the energy Hessian $\rmd^2U_i/\rmd\bfx^2$ itself, and through the scaling of the rows of the strain Jacobian $\bfS=\rmd\bfs/\rmd\bfx$.
For a regular mesh with average edge length $l$, the energy Hessian scales as $l^{-2}$ for stretching modes but $l^{-4}$ for bending, and the strain Jacobian scales as $l^{-1}$ and $l^{-2}$ for stretching and bending respectively.
We observe that multiplying ${\bf II}$ by a factor of $l^2$ compensates for both effects, improving the relative scaling of bending modes versus stretching.
In our experiments, we simply used $10^{-6}{\bf II}$ because our edge lengths were on the order of $10^{-3}$.
Second, to ensure that Newton's method results in a descent direction, the objective Hessian $\bfH$ must be modified to be positive definite.
Since eigendecomposition of $6\times6$ matrices is not readily available on the GPU, we first perform 5 iterations of the QR algorithm on $\bfH$ to bring it closer to diagonal, then estimate the smallest eigenvalue using Gershgorin bounds,
\begin{align}
\lambda_{\min} \ge \min_i \bigl(h_{ii}-\sum_{j\ne i}|h_{ij}|\bigr).
\end{align}
If the estimated $\lambda_{\min}$ is negative, we add $|\lambda_{\min}|\bfI$ to the original Hessian.

\section{Results}
\label{sec:results}

\subsection{Implementation Details}
We have implemented our GPBD algorithm in a combination of CUDA, C++, and Python.
The GPBD updates for the energy terms are performed using CUDA kernels on the GPU. We use single-precision floating point for all the operations in the neo-Hookean model, but use double-precision floating point for HYLC, because we found the Mathematica-generated energy functions of HYLC in the \cite{sperl2020hylc} implementation to be stable only when used with double precision.
For collision handling with complex obstacles, we copy the vertex positions back to the CPU and perform SDF-based projection after each GPBD iteration.
Other operations such as the initial graph colouring, computation of $\tilde\bfx$ and $\bfv$, velocity damping, and geometric constraint projections are also performed on the CPU.
Performance measurements were taken on a machine with a 13th Gen Intel(R) Core(TM) i7-13700 CPU, 64 GB RAM and an Nvidia RTX 4080 GPU. Since we use an RTX 4080 GPU, we expect to take a major performance hit for double precision operations due to 1:64 distribution of fp32 and fp64 units on the GPU. We observed a speedup of 10-15x on average when we tried to run neo-Hookean demos using single precision vs. double precision.


While the CPU-based operations and CPU-GPU memory transfer cause additional overhead in our implementation, it is not the majority of the computational cost for our examples.
The CUDA kernels take roughly 80\%-90\% of the total computation time for a $64\times64$ cloth grid, and 50\%-75\% of the total for a $128\times128$ grid.
For larger resolutions, the cost of the single-threaded CPU part of our implementation starts to grow.
Nevertheless, this is not a fundamental limitation of our approach, as the other operations could also be parallelized on the GPU.
We report computation timings only for the CUDA kernels, since that is the main compute bottleneck.
Statistics for all our examples are given in Table \ref{tab:hylc-stats}.

\begin{table}[t]
  \centering
  \begin{tabular}{llcccccc}
    \toprule
     & & \multicolumn{3}{c}{$64\times64$ grid} & \multicolumn{3}{c}{$128\times128$ grid} \\
    \cmidrule(lr){3-5}
    \cmidrule(lr){6-8}
    Scene & Material  & $i_{\rm G}$ & $i_{\rm N}$ & $t_{\rm avg}$ (ms) & $i_{\rm G}$ & $i_{\rm N}$ & $t_{\rm avg}$ (ms) \\
    \midrule
    Drape X/Y & Satin & 3 & 1 & 18/17 & 3 & 1 & 70/71  \\
    & Basket & 3 & 1 & 17/17 & 3 & 1 & 69/69  \\
    & Rib & 3 & 1 & 17/18 & 3 & 1 & 71/72  \\
    & Honey & 3 & 1 & 17/18 & 5 & 1 & 123/132  \\
    & Stock & 3 & 1 & 18/18 & 3 & 2 & 208/215  \\
    \midrule
    Stretch X/Y & Satin & 3 & 1 & 17/17 & 3 & 1 & 62/61 \\
    & Basket & 3 & 1 & 17/17 & 3 & 1 & 61/61  \\
    & Rib & 3 & 1 & 17/18 & 3 & 1 & 68/72 \\
    & Honey & 3 & 1 & 17/61 & 5 & 3 & 353/457 \\
    & Stock & 3 & 1 & 19/17 & 3 & 5 & 767/757 \\
    \midrule
    Trampoline & Satin & 3 & 1 & 377 & 3 & 1 & 70 \\
    & Basket & 3 & 1 & 373 & 3 & 1 & 69 \\
    & Rib & 3 & 1 & 375 & 3 & 1 & 71 \\
    & Honey & 3 & 1 & 402 & 3 & 3 & 222 \\
    & Stock & 3 & 1 & 384 & 3 & 3 & 282 \\
    \midrule
    Flag & Satin & 3 & 1 & 389 & 3 & 1 & 74  \\
    & Basket & 3 & 1 & 370 & 3 & 1 & 70  \\
    & Rib & 3 & 1 & 584 & 5 & 1 & 150  \\
    & Honey & 5 & 1 & 626 & 15 & 3 & 1019  \\
    & Stock & 3 & 1 & 395 & 5 & 3 & 523  \\
    \bottomrule
    
  \end{tabular}
  \caption{Statistics for HYLC examples involving square cloth run using GPBD with a timestep of $10^{-4}\,\rm s$. $i_{\rm G}$: number of GPBD iterations, $i_{\rm N}$: maximum number of Newton iterations per GPBD iteration, $t_{\rm avg}$: average computation time per timestep. The $64\times64$ grid mesh consists of 4225 vertices and 8192 triangles, while the $128\times128$ grid mesh consists of 16641 vertices and 32768 triangles.}
  \label{tab:hylc-stats}
\end{table}


\begin{table*}[t]
  \centering
  \begin{tabular}{llccccc}
    \toprule
    Scene & Material & Timestep (s) & \# verts & \# elements & GPBD/Newton iters & Avg.\ compute per timestep (ms) \\
    \midrule
    Vest on character & Stock & $10^{-4}$ & 9986 & 19580 & 3/1 & 116.9  \\
    Cube Impact & NH & $10^{-3}$ & 17576 & 93750 & 3/3 & 1.336 \\
    10\% scaled & Satin & $10^{-4}$ & 4225 & 8192 & 3/1 & 70.8  \\
    \bottomrule
  \end{tabular}
  \caption{Statistics for other examples run using GPBD.}
  \label{tab:hylc-stats-extra}
\end{table*}

\subsection{Comparison with \cite{sperl2020hylc}}

\begin{figure}
    \centering
    \includegraphics[width=\linewidth]{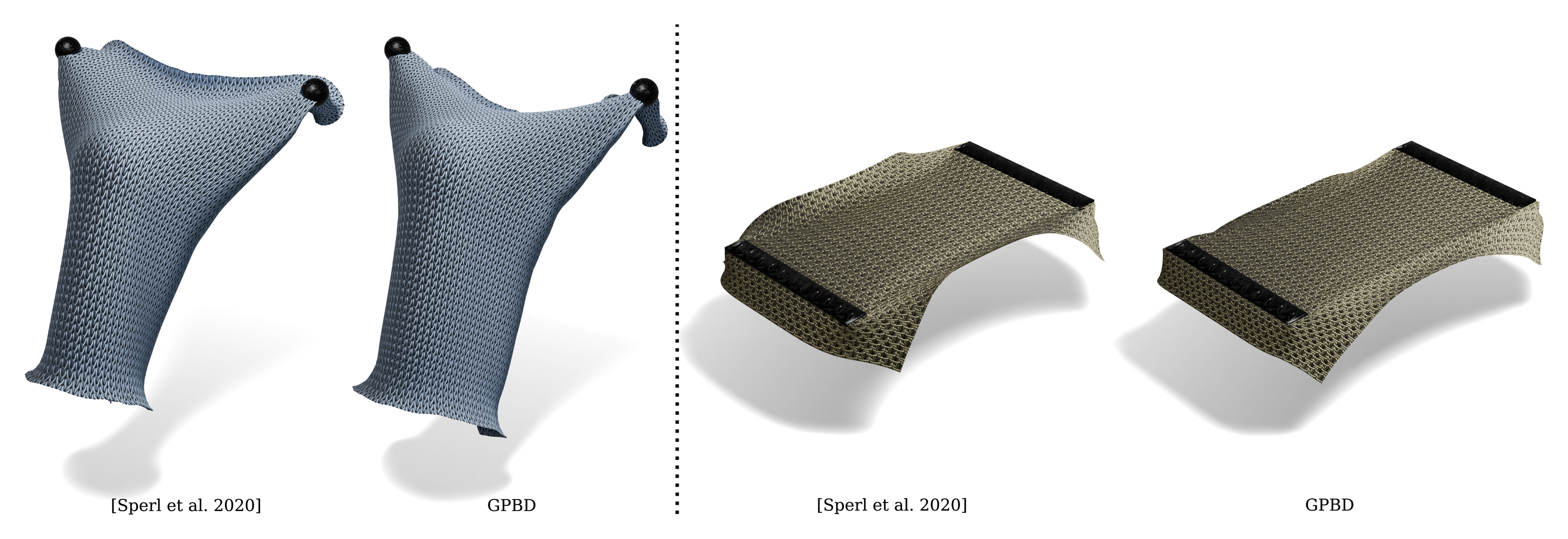}
    \caption{Qualitative comparison between our method GPBD and \cite{sperl2020hylc} on HYLC material models: Stockinette drape (left) and Honeycomb stretch (right).}
    \label{fig:hylc-comparison}
\end{figure}

\begin{figure}[htbp]
    \begin{minipage}{0.36\textwidth}
        \centering
        \includegraphics[width=\linewidth]{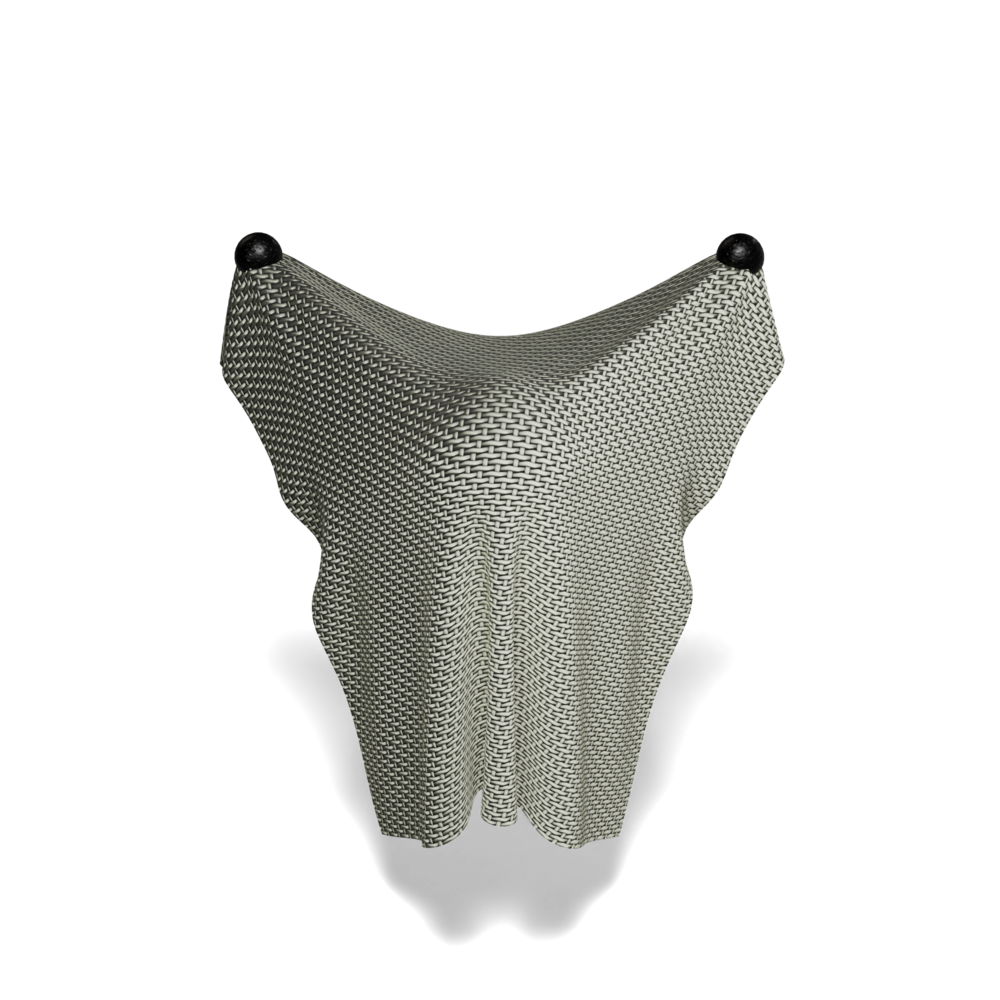}
        \caption{10\% scaled Satin material $128 \times 128$ grid mesh (16641 verts, 32768 tris) with wrinkles.}
        \label{fig:hylc-scaled}
    \end{minipage}%
    \hfill%
    \begin{minipage}{0.50\textwidth}
        \centering
        \includegraphics[width=\linewidth]{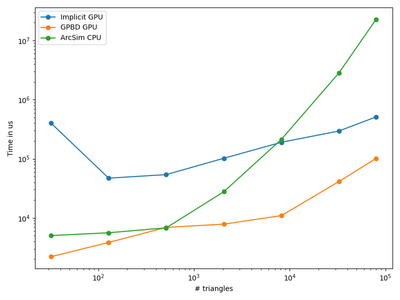}
        \caption{A log-log plot of the average frame time vs. mesh resolution for different simulation methods. Timings are averaged over 100 frames with a timestep of $10^{-4}\,\mathrm s$ for a cloth sheet with the HYLC Satin material model stretched from two sides.}
        \label{fig:hylc-perf=comparison}
    \end{minipage}
\end{figure}

\begin{figure*}[h]

  \centering
  \includegraphics[width=\linewidth]{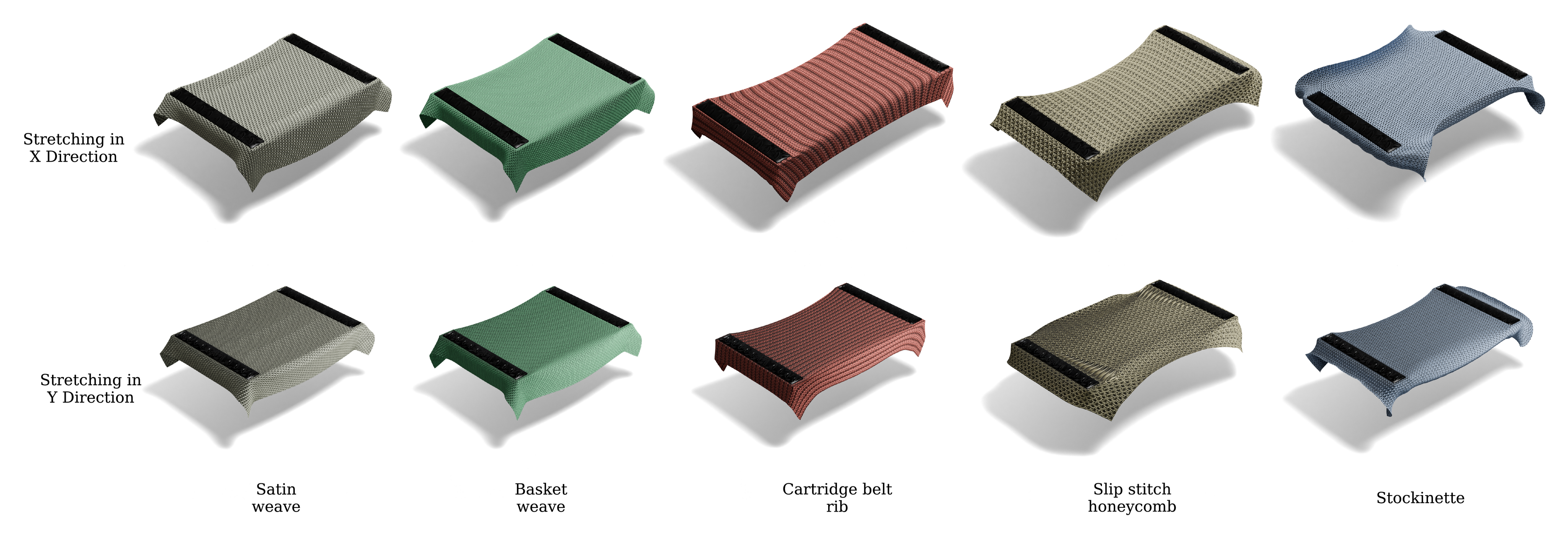}
  \caption{"Cloth stretched". $128\times128$ grid mesh(16641 verts, 32768 tris) results on data-driven cloth material. The reader is encouraged to compare these with \cite{sperl2020hylc} Fig 16}
  \label{fig:hylc-stretch}

  \centering
    \includegraphics[width=\linewidth]{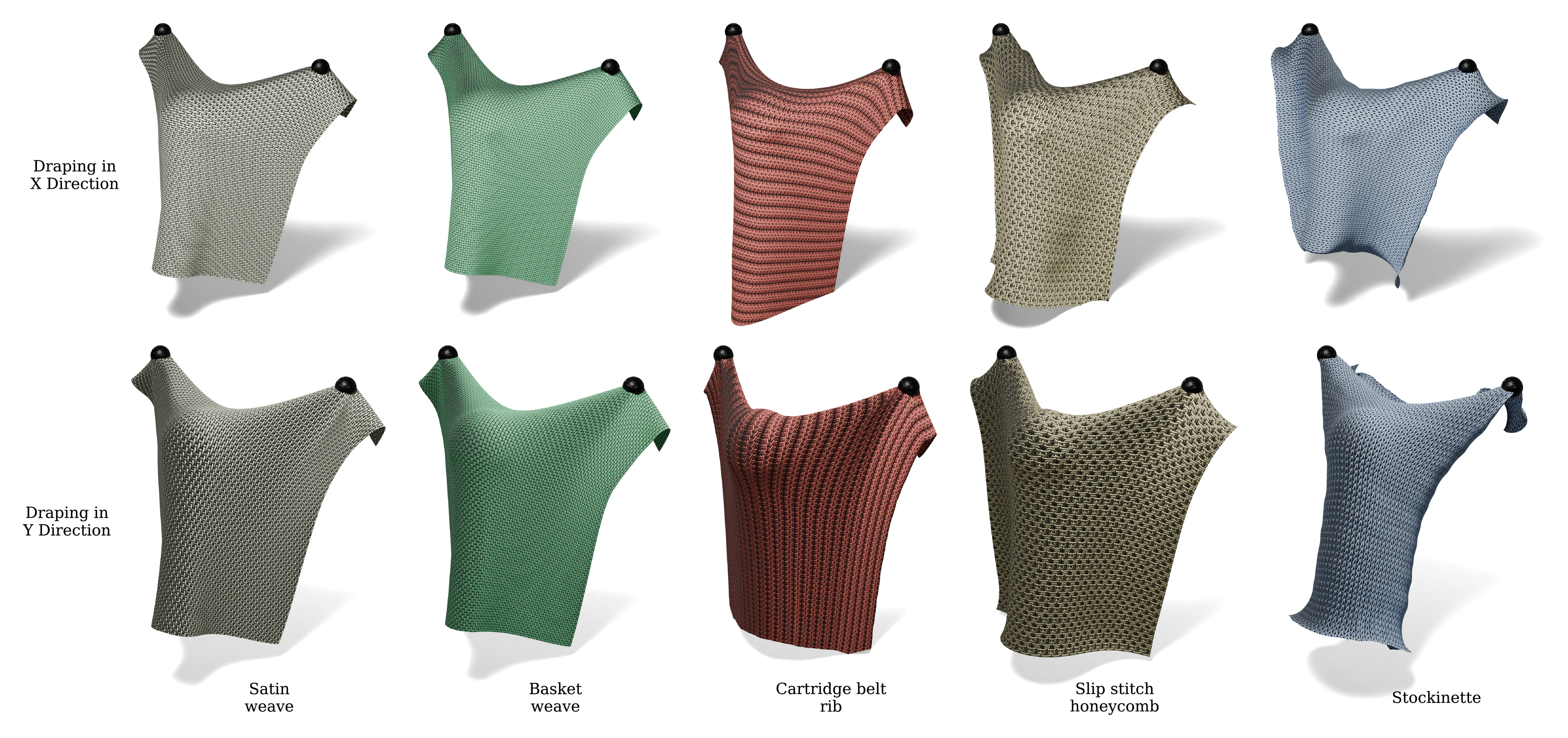}
    \caption{Cloth draped on sphere. $128 \times 128$ grid mesh (16641 verts, 32768 tris) results on data-driven cloth material. The reader is encouraged to compare these with \cite{sperl2020hylc} Fig 16.}
    \label{fig:hylc-drape}
\end{figure*}

To validate the fidelity of our simulations, we run the same draping and stretching examples from the original HYLC implementation, with all five fabrics: Satin, Basket weave, Honeycomb, Rib knit, and Stockinette.
The results are shown in Figures~\ref{fig:hylc-drape} and \ref{fig:hylc-stretch} respectively.
Many of the fabrics, especially Rib knit and Stockinette, are highly anisotropic and exhibit different stretching and curling behaviours in different directions.
As can be seen in the \autoref{fig:hylc-comparison}, our simulation faithfully reproduces this behaviour, and closely matches the results of the original HYLC implementation \cite{sperl2020hylc} based on the ARCSim cloth simulator. 
For a fair comparison with our method, we disabled remeshing and self-collision handling in the original HYLC implementation.
We encourage the reader to compare our results in the supplementary document. 
In addition, we show an example of the satin material at 0.1x scale in Figure~\ref{fig:hylc-scaled}, exhibiting finer wrinkles.


The computational cost of our method is compared with \citet{sperl2020hylc}'s original ARCSim implementation in 
Figure ~\ref{fig:hylc-perf=comparison}. 
To disentangle the performance benefits of GPBD vs.~GPU parallelization, we also implemented a global solver for backward Euler on the GPU using Nvidia's cuDSS (direct sparse solver) library, which we refer to as Implicit GPU.
As shown, the original CPU's runtime increases rapidly with mesh resolution since it performs implicit time integration using a global Newton solve on the entire system.
In contrast, both GPBD GPU and Implicit GPU exhibit more moderate scaling. 

To further understand our iteration bottlenecks for HYLC energy model, we ran the simulation on CPU for a single element with 1 GPBD iteration and 1 Newton iteration. Timings for various parts of the algorithm for a single time step are listed in Table~\ref{tab:breakdown}.
The force and force Jacobian computations take up the majority of the computation time. Inside line search, the majority of compute time is taken up by the repeated evaluation of the material energy function. Hence, the main bottlenecks in our code arise from the Mathematica-generated energy functions and their derivatives from the \cite{sperl2020hylc} implementation.
A more optimized implementation of the material energy would provide an immediate speedup without any changes to our algorithm.

\begin{table}[h]
\centering
\begin{tabular}{lcc}
    \toprule
    \multicolumn{1}{c}{\textbf{Step}} & \multicolumn{1}{c}{\textbf{Time(ms)}} & \multicolumn{1}{c}{\textbf{\% of total time}} \\
    \midrule
    Force \& force Jacobian eval & 71.864 & 57.67 \\
    Linear solve & 4.53 & 3.63 \\
    Line search & 47.89 & 38.43 \\
    \bottomrule
\end{tabular}
\caption{Time breakdown of different steps in the algorithm.}
\label{tab:breakdown}
\end{table}

Finally, we point out that most of our simulations only require a small number of GPBD iterations per time step.
However, as noted in Table~\ref{tab:hylc-stats}, some examples involving strongly nonlinear materials under large deformation require more iterations for stability.
In the stretching demo, for the Stockinette material on the $128\times128$ grid, we observed jittering even at large iteration counts.
Therefore, in the figures, we show results on a $128 \times 128$ grid for all materials except for Stockinette in the stretching demo, for which the $64\times64$ grid is shown.

\subsection{Neo-Hookean elasticity}

\begin{figure}
    \centering
    \includegraphics[width=\linewidth]{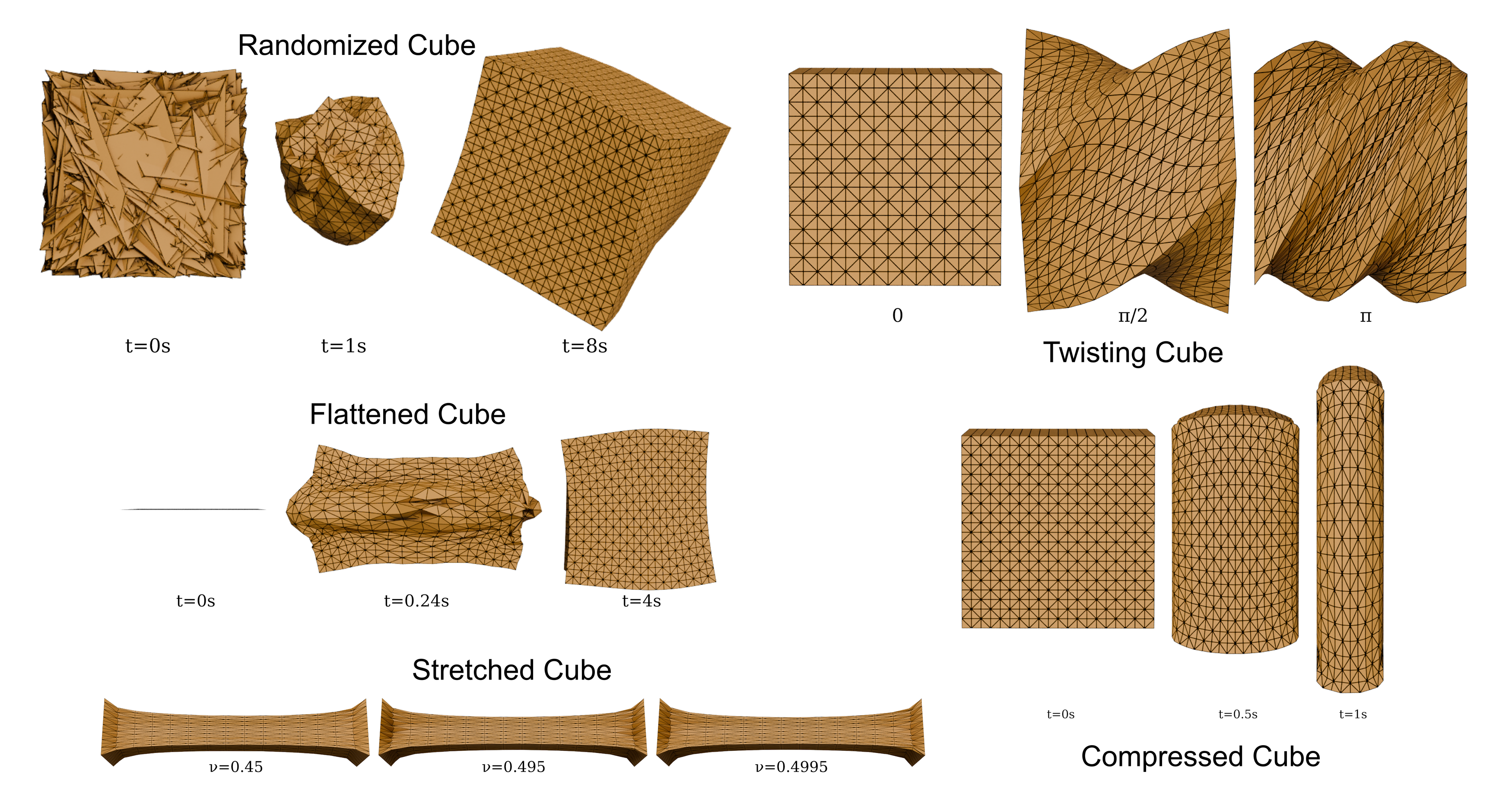}
    \caption{We perform various stress tests on a neo-Hookean energy model using a cube composed of $15^3$ and $20^3$ hexahedral cells.}
    \label{fig:volumetric-demos}
\end{figure}

We show several examples in Figure~\ref{fig:volumetric-demos} and Table~\ref{tab:nh-stats} that demonstrate stable simulations with high Poisson ratios. We attempt to replicate the demos of \citet{macklin2021neohookean} with both the log-barrier neo-Hookean model \citep{sifakis2012fem} and the stable neo-Hookean model \citep{smith2018flesh}. We are able to obtain stable simulations for all examples at $0.001\,\rms$ timestep, except for the stable neo-Hookean model at a Poisson's ratio of 0.4995 --- surprisingly, the log-barrier variant of neo-Hookean remains stable at the same Poisson's ratio.
All our other demos can be handled stably with up to 3 GPBD iterations and 20 max Newton iterations. Since our algorithm performs an early exit if the Newton solver converges earlier, increasing the max Newton iterations to $20$ does not incur much of a performance hit. In our experiments, this helps the solver to deal with energy domain where the log barrier causes the optimization problem to become highly non-linear, thus requiring many Newton iterations to prevent artifacts and instabilities.

\begin{figure}
    \centering
    \includegraphics[width=\linewidth]{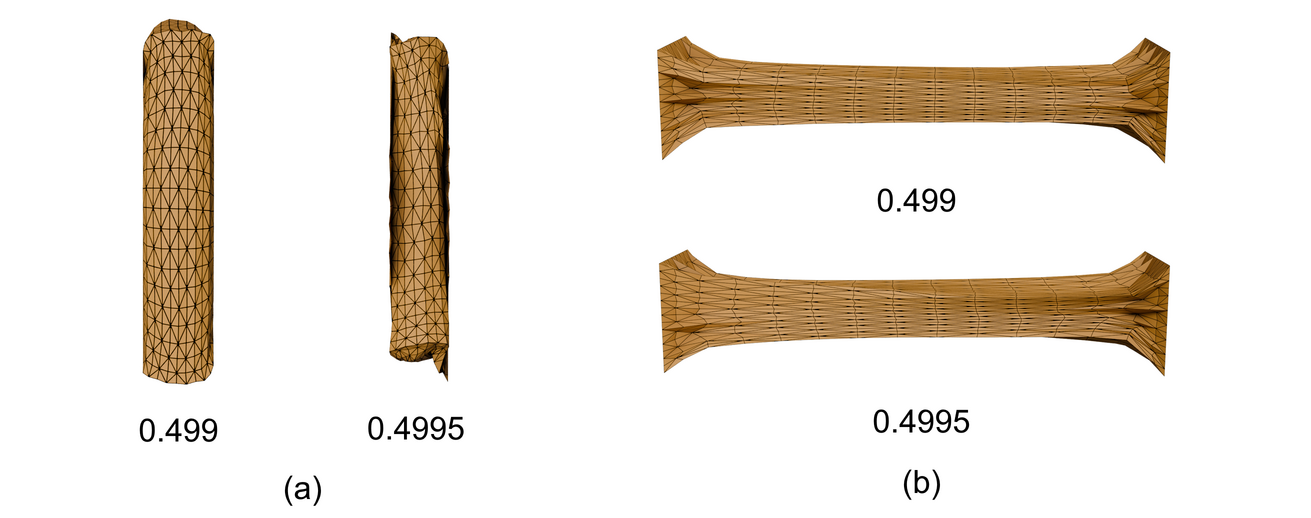}
    \caption{The same tests were carried out using 64 VBD iterations per frame for SNH cubes. We observe that we get incorrect results for higher Poisson ratios.}
    \label{fig:vbd-failure}
\end{figure}

We also compare our method with the recent vertex block descent (VBD) method of \citet{chen2024vbd} on the stable neo-Hookean model. 
The per-iteration cost of both methods is shown in Table~\ref{tab:vbd}.
VBD performs Gauss-Seidel updates on vertices, requiring fewer graph colours than per-element Gauss-Seidel.
However, the Jacobi version of GPBD does not require any graph colouring, and offers better performance than VBD.
Theoretically, VBD has the advantage of offering unconditional stability, since with line search it can be ensured to always performs descent steps; however, this significantly degrades performance and the authors argue for leaving line search disabled.
We observe that without line search, the results of VBD sometimes exhibit objectionable artifacts under high-stress scenarios.
Some such artifacts are shown in Figure~\ref{fig:vbd-failure}.
Finally, we also note that VBD requires additional care to handle materials with an inversion barrier, as the presence of an inverted or degenerate element will render adjacent vertices unable to be updated.
Only if the mesh is initialized in a fully non-inverted state, and line search is always enabled, will it be possible to ensure simulation progress.
By contrast, as shown in Figure~\ref{fig:volumetric-demos}, our method supports NH with an inversion barrier even when starting from highly degraded initializations.


\begin{table}[h]
\centering
\begin{tabular}{lc}
    \toprule
    \multicolumn{1}{c}{\textbf{Method}} & \multicolumn{1}{c}{\textbf{Average frame time ($\mu$s)}}\\
    \midrule
    VBD & 113 \\
    GPBD Gauss seidel & 629 \\
    GPBD Jacobi & 102 \\
    GPBD Gauss seidel theoretical limit & 357 \\
    GPBD Jacobi theoretical limit & 49 \\
    \bottomrule
\end{tabular}
\caption{Average frame time for simulating a $25\times 25\times 25$ SNH cube without backtracking}
\label{tab:vbd}
\end{table}


\subsection{Additional examples}

\begin{figure*}[h]
  \centering
  \includegraphics[width=\linewidth]{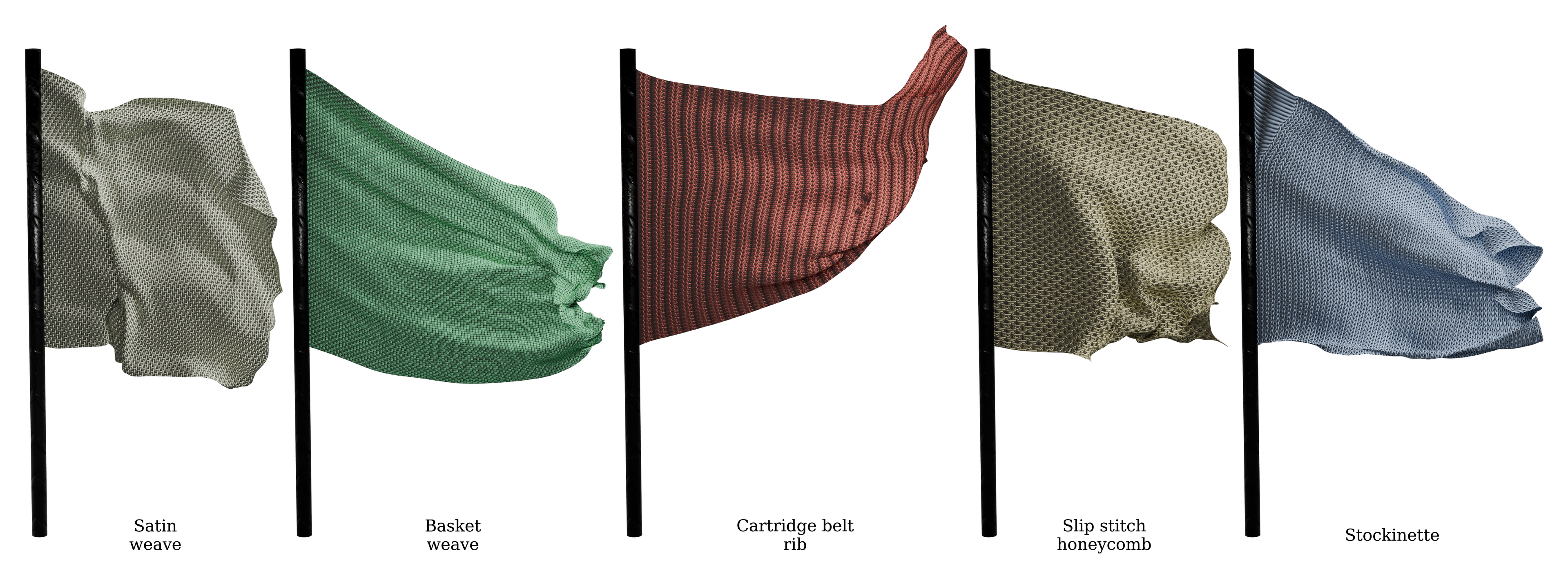}
  \caption{"Flag in wind". A $128\times128$ grid cloth mesh(16641 verts, 32768 tris) with fixed corners, attached to a flagpole in simulated wind.} 
  \label{fig:flags}
\end{figure*}

In Figure~\ref{fig:flags} we simulate flags of different materials waving in a constant wind field.
Wind is modeled as an external force applying lift and drag forces on the cloth surface.
Different materials exhibit significantly different behaviour under the same wind field.

\begin{figure*}[h]
  \centering
  \includegraphics[width=\linewidth]{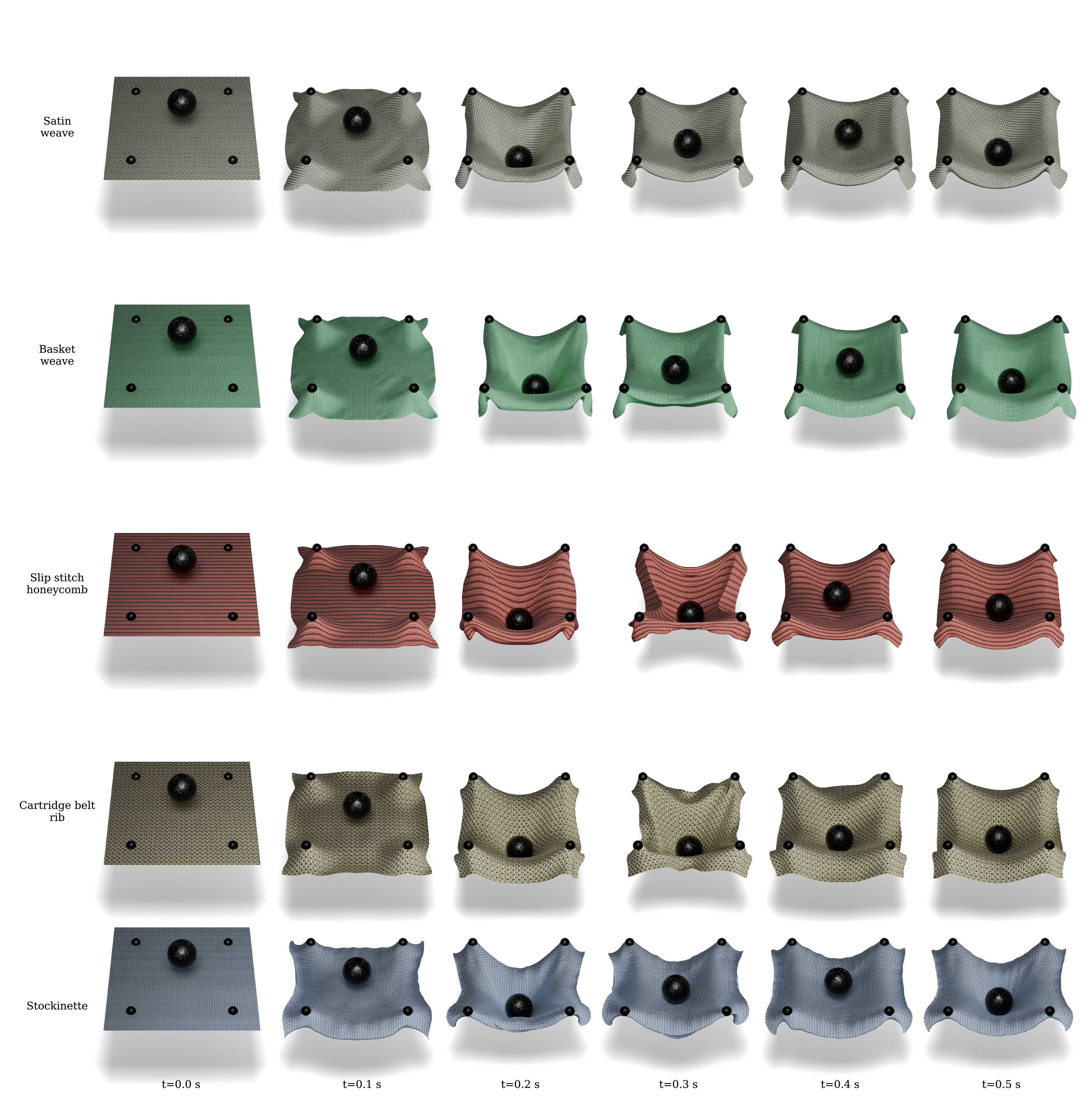}
  \caption{"Ball dropped on a trampoline". A light ball dropped on top of a $128\times128$ grid cloth mesh(16641 verts, 32768 tris) with fixed corners.}
  \label{fig:trampoline}
\end{figure*}

As discussed in Section~\ref{sec:gpbd}, it is straightforward to combine our method with existing PBD approaches.
We show two-way collision handling with a rigid sphere in Figure~\ref{fig:trampoline}.
Collisions between the cloth particles and the sphere are modeled as inequality constraints, and resolved simply by projecting the particles and the sphere away from each other.
The same example is shown with different materials in the supplementary video.

Similarly, collisions with more complex obstacles can be handled by projection.
A more complex example of a Stockinette vest \cite{bertiche2020cloth3d} on an animated character \cite{SMPL-X:2019} is shown in Figure~\ref{fig:teaser} (left).
As expected, the characteristic curling behaviour of the Stockinette material is visible on the edges of the garment.



\begin{table*}[t]
  \centering
  \begin{tabular}{lcccccccc}
    \toprule
    Scene & \#Verts & \#Tets & Poisson ratio & Timestep (s) & $i_{\rm G}$ & $i_{\rm N}$ & Sim length (s) & $t_{\rm avg} (\mu\rm)$ \\
    \midrule
    Flatten NH & 9261 & 48000 & 0.4995 & 0.01 & 2 & 6 & 20 & 399 \\
    Randomize NH & 9261 & 48000 & 0.4995 & 0.01 & 2 & 8 & 20 & 1014 \\
    Stretch SNH/NH & 4096 & 20250 & 0.45 & 0.001 & 3/3 & 2/3 & 2 & 257/450\\
    & & & 0.49 & 0.001 & 2/3 & 3/5 & 2 & 251/731\\
    & & & 0.495 & 0.001 & 2/3 & 3/5 & 2 & 250/737\\
    & & & 0.499 & 0.001 & 3/3 & 5/5 & 2 & 569/730\\
    & & & 0.4995 & 0.001 & -/3 & -/5 & 2 & -/1.450\\
    Compress SNH/NH & 9261 & 48000 & 0.45 & 0.001 & 2/2 & 3/5 & 2 & 325/759\\
    & & & 0.49 & 0.001 & 2/2 & 3/7 & 2 & 322/1056\\
    & & & 0.495 & 0.001 & 2/2 & 3/10 & 2 & 322/1382\\
    & & & 0.499 & 0.001 & 2/8 & 5/10 & 2 & 628/5398\\
    & & & 0.4995 & 0.001 & 2/20 & 8/10 & 2 & 984/13437\\
    Twist $\pi$ SNH/NH & 4096 & 20250 & 0.45 & 0.001 & 2/2 & 1/2 & 2 & 132/273\\
    & & & 0.49 & 0.001 & 2/2 & 1/2 & 2 & 134/274\\
    & & & 0.495 & 0.001 & 2/2 & 2/2 & 2 & 184/267\\
    & & & 0.499 & 0.001 & 2/2 & 2/5 & 2 & 184/505\\
    & & & 0.4995 & 0.001 & 2/2 & 3/5 & 2 & 236/507\\
    \bottomrule
  \end{tabular}
  \caption{Statistics for stable neo-Hookean and neo-Hookean examples executed using GPBD on a  tessellated cube, with Young's modulus of $10^5$. $i_{\rm G}$: number of GPBD iterations, $i_{\rm N}$: maximum number of Newton iterations per GPBD iteration, $t_{\rm avg}$: average simulation time per timestep. }
  \label{tab:nh-stats}
\end{table*}

\section{Conclusion}

We have presented a technique for fast simulation of nonlinear deformable models, based on expressing the equations of implicit time integration in terms of the displacements caused by all the forces acting on the system.
Our approach generalizes the position-based dynamics framework beyond the traditional constraint-based formulation, allowing it to support arbitrary nonlinear material models.
This enables scalable simulation of complex materials like neo-Hookean and data-driven cloth.

\subsection{Discussion and Limitations}

The FP-PXPBD method proposed by \citet{chen2023hyperelasticity} is closely related to our algorithm.
Similar to our method, their approach solves the update equations for one element at a time, holding the forces due to all adjacent elements fixed.
However, their approach is tied specifically to hyperelastic finite elements, while our method applies to general discretizations.

Unfortunately, our algorithm is not unconditionally stable, and requires sufficiently many GPBD iterations to produce a stable simulation.
For most of our examples, 3 GPBD iterations are enough to give good results, but some materials required a larger number of iterations to avoid jittering.

\citet{tonthat2023parallel} have reported similar instabilities in the XPBD simulation of neo-Hookean finite elements \cite{macklin2021neohookean}.
Hence we feel that the stability properties of position-based methods are not fully understood and deserve further study.

For the neo-Hookean model, we begin to observe oscillations in the simulation for Poisson's ratios of $0.499$ and above, because even the local optimization problem becomes ill-conditioned and our Newton solver exhibits poorer convergence. 
For the HYLC model, the computational bottleneck is the evaluation of the material energy and its gradient and Hessian, rather than the Newton solve involved in the GPBD update.
Making it stable in single precision floating point and replacing it with a faster-to-evaluate model, such as the neural HYLC approach of \citet{feng2024neural}, would give an immediate speedup.
Our experiments showed that instabilities often start with a local element update leading to an overshoot in the global optimisation objective, \cite{10.1145/3731183} presents a promising formulation which can help with incorporating global information in local GPBD updates and avoid this overshoot.


Finally, we point out one significant opportunity enabled by our GPBD formulaion: we can observe that the GPBD update problem \eqref{eq:gpbd-reduced-opt} is reminiscent of a proximal operator \cite{parikh2014proximal} applied to the energy $U_i$ at the current strain $\bfs_i$.
This opens the door to precomputing the solution operator $\bfs_i \mapsto \Delta\bflambda_i$, perhaps by learning it with a small neural network.
This solution operator could then immediately be evaluated at runtime to obtain $\Delta\bflambda_i$ in constant time, without performing any Newton iterations. 
We are looking forward to exploring this approach in future work.
According to our analysis, it could reduce the run-time cost by up to a factor of 2 (if the cost of the solver becomes negligible).

\begin{acks}
We thank Argha Chakraborty and Abhay Pratap Singh Rathore for help with initial implementation of this work. This work was partially supported by ANRF grant CRG/2023/009042.
\end{acks}


\bibliographystyle{ACM-Reference-Format}
\bibliography{bibliography}

@article{bender2014continuous,
title = {Position-based simulation of continuous materials},
journal = {Computers \& Graphics},
volume = {44},
pages = {1-10},
year = {2014},
issn = {0097-8493},
doi = {https://doi.org/10.1016/j.cag.2014.07.004},
url = {https://www.sciencedirect.com/science/article/pii/S009784931400065X},
author = {Jan Bender and Dan Koschier and Patrick Charrier and Daniel Weber},
}

@inproceedings {bender2017survey,
booktitle = {EG 2017 - Tutorials},
editor = {Adrien Bousseau and Diego Gutierrez},
title = {{A Survey on Position Based Dynamics}},
author = {Bender, Jan and Müller, Matthias and Macklin, Miles},
year = {2017},
publisher = {The Eurographics Association},
ISSN = {1017-4656},
DOI = {10.2312/egt.20171034}
}

@article{bouaziz2014projective,
author = {Bouaziz, Sofien and Martin, Sebastian and Liu, Tiantian and Kavan, Ladislav and Pauly, Mark},
title = {Projective dynamics: fusing constraint projections for fast simulation},
year = {2014},
issue_date = {July 2014},
publisher = {Association for Computing Machinery},
address = {New York, NY, USA},
volume = {33},
number = {4},
issn = {0730-0301},
url = {https://doi.org/10.1145/2601097.2601116},
doi = {10.1145/2601097.2601116},
journal = {ACM Trans. Graph.},
month = {jul},
articleno = {154},
numpages = {11}
}

@inproceedings{chen2023hyperelasticity,
author = {Chen, Yizhou and Han, Yushan and Chen, Jingyu and Ma, Shiqian and Fedkiw, Ronald and Teran, Joseph},
title = {Primal Extended Position Based Dynamics for Hyperelasticity},
year = {2023},
isbn = {9798400703935},
publisher = {Association for Computing Machinery},
address = {New York, NY, USA},
url = {https://doi.org/10.1145/3623264.3624437},
doi = {10.1145/3623264.3624437},
booktitle = {Proceedings of the 16th ACM SIGGRAPH Conference on Motion, Interaction and Games},
articleno = {21},
numpages = {10},
location = {<conf-loc>, <city>Rennes</city>, <country>France</country>, </conf-loc>},
series = {MIG '23}
}

@misc{chen2024vbd,
      title={Vertex Block Descent}, 
      author={Anka He Chen and Ziheng Liu and Yin Yang and Cem Yuksel},
      year={2024},
      eprint={2403.06321},
      archivePrefix={arXiv},
      primaryClass={cs.GR}
}

@article{deul2014rigid,
author = {Deul, Crispin and Charrier, Patrick and Bender, Jan},
title = {Position-based rigid-body dynamics},
journal = {Computer Animation and Virtual Worlds},
volume = {27},
number = {2},
pages = {103-112},
doi = {https://doi.org/10.1002/cav.1614},
url = {https://onlinelibrary.wiley.com/doi/abs/10.1002/cav.1614},
eprint = {https://onlinelibrary.wiley.com/doi/pdf/10.1002/cav.1614},
year = {2014}
}

@inproceedings{diziol2011incompressible,
author = {Diziol, R. and Bender, J. and Bayer, D.},
title = {Robust Real-Time Deformation of Incompressible Surface Meshes},
year = {2011},
isbn = {9781450309233},
publisher = {Association for Computing Machinery},
address = {New York, NY, USA},
url = {https://doi.org/10.1145/2019406.2019438},
doi = {10.1145/2019406.2019438},
booktitle = {Proceedings of the 2011 ACM SIGGRAPH/Eurographics Symposium on Computer Animation},
pages = {237–246},
numpages = {10},
location = {Vancouver, British Columbia, Canada},
series = {SCA '11}
}

@inproceedings{feng2024neural,
author = {Feng, Xudong and Wang, Huamin and Yang, Yin and Xu, Weiwei},
title = {Neural-Assisted Homogenization of Yarn-Level Cloth},
year = {2024},
isbn = {9798400705250},
publisher = {Association for Computing Machinery},
address = {New York, NY, USA},
url = {https://doi.org/10.1145/3641519.3657411},
doi = {10.1145/3641519.3657411},
booktitle = {ACM SIGGRAPH 2024 Conference Papers},
articleno = {80},
numpages = {10},
location = {Denver, CO, USA},
series = {SIGGRAPH '24}
}

@article{fratarcangeli2016vivace,
 author = {Fratarcangeli, Marco and Tibaldo, Valentina and Pellacini, Fabio},
 title = {Vivace: A Practical Gauss-seidel Method for Stable Soft Body Dynamics},
 journal = {ACM Trans. Graph.},
 issue_date = {November 2016},
 volume = {35},
 number = {6},
 month = nov,
 year = {2016},
 issn = {0730-0301},
 pages = {214:1--214:9},
 articleno = {214},
 numpages = {9},
 url = {http://doi.acm.org/10.1145/2980179.2982437},
 doi = {10.1145/2980179.2982437},
 acmid = {2982437},
 publisher = {ACM},
 address = {New York, NY, USA},
}

@inproceedings {kugelstadt2016rods,
booktitle = {Eurographics/ ACM SIGGRAPH Symposium on Computer Animation},
editor = {Ladislav Kavan and Chris Wojtan},
title = {{Position and Orientation Based Cosserat Rods}},
author = {Kugelstadt, Tassilo and Schömer, Elmar},
year = {2016},
publisher = {The Eurographics Association},
ISSN = {1727-5288},
ISBN = {978-3-03868-009-3},
DOI = {10.2312/sca.20161234}
}

@article{lan2023stencil,
author = {Lan, Lei and Li, Minchen and Jiang, Chenfanfu and Wang, Huamin and Yang, Yin},
title = {Second-order Stencil Descent for Interior-point Hyperelasticity},
year = {2023},
issue_date = {August 2023},
publisher = {Association for Computing Machinery},
address = {New York, NY, USA},
volume = {42},
number = {4},
issn = {0730-0301},
url = {https://doi.org/10.1145/3592104},
doi = {10.1145/3592104},
journal = {ACM Trans. Graph.},
month = {jul},
articleno = {108},
numpages = {16}
}

@article{liu2017quasinewton,
author = {Liu, Tiantian and Bouaziz, Sofien and Kavan, Ladislav},
title = {Quasi-Newton Methods for Real-Time Simulation of Hyperelastic Materials},
year = {2017},
issue_date = {June 2017},
publisher = {Association for Computing Machinery},
address = {New York, NY, USA},
volume = {36},
number = {3},
issn = {0730-0301},
url = {https://doi.org/10.1145/2990496},
doi = {10.1145/2990496},
journal = {ACM Trans. Graph.},
month = {may},
articleno = {23},
numpages = {16}
}

@article{macklin2013fluids,
author = {Macklin, Miles and M\"{u}ller, Matthias},
title = {Position Based Fluids},
year = {2013},
issue_date = {July 2013},
publisher = {Association for Computing Machinery},
address = {New York, NY, USA},
volume = {32},
number = {4},
issn = {0730-0301},
url = {https://doi.org/10.1145/2461912.2461984},
doi = {10.1145/2461912.2461984},
journal = {ACM Trans. Graph.},
month = {jul},
articleno = {104},
numpages = {12}
}

@article{macklin2014unified,
author = {Macklin, Miles and M\"{u}ller, Matthias and Chentanez, Nuttapong and Kim, Tae-Yong},
title = {Unified Particle Physics for Real-Time Applications},
year = {2014},
issue_date = {July 2014},
publisher = {Association for Computing Machinery},
address = {New York, NY, USA},
volume = {33},
number = {4},
issn = {0730-0301},
url = {https://doi.org/10.1145/2601097.2601152},
doi = {10.1145/2601097.2601152},
journal = {ACM Trans. Graph.},
month = {jul},
articleno = {153},
numpages = {12}
}

@inproceedings{macklin2016xpbd,
author = {Macklin, Miles and M\"{u}ller, Matthias and Chentanez, Nuttapong},
title = {XPBD: Position-Based Simulation of Compliant Constrained Dynamics},
year = {2016},
isbn = {9781450345927},
publisher = {Association for Computing Machinery},
address = {New York, NY, USA},
url = {https://doi.org/10.1145/2994258.2994272},
doi = {10.1145/2994258.2994272},
booktitle = {Proceedings of the 9th International Conference on Motion in Games},
pages = {49–54},
numpages = {6},
location = {Burlingame, California},
series = {MIG '16}
}

@inproceedings{macklin2019small,
author = {Macklin, Miles and Storey, Kier and Lu, Michelle and Terdiman, Pierre and Chentanez, Nuttapong and Jeschke, Stefan and M\"{u}ller, Matthias},
title = {Small Steps in Physics Simulation},
year = {2019},
isbn = {9781450366779},
publisher = {Association for Computing Machinery},
address = {New York, NY, USA},
url = {https://doi.org/10.1145/3309486.3340247},
doi = {10.1145/3309486.3340247},
booktitle = {Proceedings of the 18th Annual ACM SIGGRAPH/Eurographics Symposium on Computer Animation},
articleno = {2},
numpages = {7},
location = {Los Angeles, California},
series = {SCA '19}
}

@article{macklin2020pd,
author = {Macklin, M. and Erleben, K. and Müller, M. and Chentanez, N. and Jeschke, S. and Kim, T.Y.},
title = {Primal/Dual Descent Methods for Dynamics},
journal = {Computer Graphics Forum},
volume = {39},
number = {8},
pages = {89-100},
doi = {https://doi.org/10.1111/cgf.14104},
url = {https://onlinelibrary.wiley.com/doi/abs/10.1111/cgf.14104},
eprint = {https://onlinelibrary.wiley.com/doi/pdf/10.1111/cgf.14104},
year = {2020}
}

@inproceedings{macklin2021neohookean,
author = {Macklin, Miles and Muller, Matthias},
title = {A Constraint-Based Formulation of Stable Neo-Hookean Materials},
year = {2021},
isbn = {9781450391313},
publisher = {Association for Computing Machinery},
address = {New York, NY, USA},
url = {https://doi.org/10.1145/3487983.3488289},
doi = {10.1145/3487983.3488289},
booktitle = {Motion, Interaction and Games},
articleno = {12},
numpages = {7},
location = {Virtual Event, Switzerland},
series = {MIG '21}
}

@inproceedings {mueller2006pbd,
booktitle = {Vriphys: 3rd Workshop in Virtual Realitiy, Interactions, and Physical Simulation},
editor = {Cesar Mendoza and Isabel Navazo},
title = {{Position Based Dynamics}},
author = {Müller, Matthias and Heidelberger, Bruno and Hennix, Marcus and Ratcliff, John},
year = {2006},
publisher = {The Eurographics Association},
ISBN = {3-905673-61-4},
DOI = {10.2312/PE/vriphys/vriphys06/071-080}
}

@inproceedings {mueller2014strain,
booktitle = {Eurographics/ ACM SIGGRAPH Symposium on Computer Animation},
editor = {Vladlen Koltun and Eftychios Sifakis},
title = {{Strain Based Dynamics}},
author = {Müller, Matthias and Chentanez, Nuttapong and Kim, Tae-Yong and Macklin, Miles},
year = {2014},
publisher = {The Eurographics Association},
ISSN = {1727-5288},
ISBN = {978-3-905674-61-3},
DOI = {10.2312/sca.20141133}
}

@article {mueller2020rigid,
journal = {Computer Graphics Forum},
title = {{Detailed Rigid Body Simulation with Extended Position Based Dynamics}},
author = {Müller, Matthias and Macklin, Miles and Chentanez, Nuttapong and Jeschke, Stefan and Kim, Tae-Yong},
year = {2020},
publisher = {The Eurographics Association and John Wiley & Sons Ltd.},
ISSN = {1467-8659},
DOI = {10.1111/cgf.14105}
}

@article{parikh2014proximal,
url = {http://dx.doi.org/10.1561/2400000003},
year = {2014},
volume = {1},
journal = {Foundations and Trends® in Optimization},
title = {Proximal Algorithms},
doi = {10.1561/2400000003},
issn = {2167-3888},
number = {3},
pages = {127-239},
author = {Neal Parikh and Stephen Boyd}
}

@article{peng2018anderson,
author = {Peng, Yue and Deng, Bailin and Zhang, Juyong and Geng, Fanyu and Qin, Wenjie and Liu, Ligang},
title = {Anderson acceleration for geometry optimization and physics simulation},
year = {2018},
issue_date = {August 2018},
publisher = {Association for Computing Machinery},
address = {New York, NY, USA},
volume = {37},
number = {4},
issn = {0730-0301},
url = {https://doi.org/10.1145/3197517.3201290},
doi = {10.1145/3197517.3201290},
journal = {ACM Trans. Graph.},
month = {jul},
articleno = {42},
numpages = {14}
}

@inproceedings{sifakis2012fem,
author = {Sifakis, Eftychios and Barbic, Jernej},
title = {FEM simulation of 3D deformable solids: a practitioner's guide to theory, discretization and model reduction},
year = {2012},
isbn = {9781450316781},
publisher = {Association for Computing Machinery},
address = {New York, NY, USA},
url = {https://doi.org/10.1145/2343483.2343501},
doi = {10.1145/2343483.2343501},
booktitle = {ACM SIGGRAPH 2012 Courses},
articleno = {20},
numpages = {50},
location = {Los Angeles, California},
series = {SIGGRAPH '12}
}

@article{smith2018flesh,
author = {Smith, Breannan and Goes, Fernando De and Kim, Theodore},
title = {Stable Neo-Hookean Flesh Simulation},
year = {2018},
issue_date = {April 2018},
publisher = {Association for Computing Machinery},
address = {New York, NY, USA},
volume = {37},
number = {2},
issn = {0730-0301},
url = {https://doi.org/10.1145/3180491},
doi = {10.1145/3180491},
month = {mar},
articleno = {12},
numpages = {15}
}

@article{sperl2020hylc,
author = {Sperl, Georg and Narain, Rahul and Wojtan, Chris},
title = {Homogenized yarn-level cloth},
year = {2020},
issue_date = {August 2020},
publisher = {Association for Computing Machinery},
address = {New York, NY, USA},
volume = {39},
number = {4},
issn = {0730-0301},
url = {https://doi.org/10.1145/3386569.3392412},
doi = {10.1145/3386569.3392412},
journal = {ACM Trans. Graph.},
month = {aug},
articleno = {48},
numpages = {16}
}

@article{tonthat2023parallel,
title = {Parallel block Neo-Hookean XPBD using graph clustering},
journal = {Computers \& Graphics},
volume = {110},
pages = {1-10},
year = {2023},
issn = {0097-8493},
doi = {https://doi.org/10.1016/j.cag.2022.10.009},
url = {https://www.sciencedirect.com/science/article/pii/S009784932200187X},
author = {Quoc-Minh Ton-That and Paul G. Kry and Sheldon Andrews}
}

@inproceedings{umetani2014rods,
author = {Umetani, Nobuyuki and Schmidt, Ryan and Stam, Jos},
title = {Position-Based Elastic Rods},
year = {2014},
publisher = {Eurographics Association},
address = {Goslar, DEU},
booktitle = {Proceedings of the ACM SIGGRAPH/Eurographics Symposium on Computer Animation},
pages = {21–30},
numpages = {10},
location = {Copenhagen, Denmark},
series = {SCA '14}
}

@article{wang2015chebyshev,
author = {Wang, Huamin},
title = {A chebyshev semi-iterative approach for accelerating projective and position-based dynamics},
year = {2015},
issue_date = {November 2015},
publisher = {Association for Computing Machinery},
address = {New York, NY, USA},
volume = {34},
number = {6},
issn = {0730-0301},
url = {https://doi.org/10.1145/2816795.2818063},
doi = {10.1145/2816795.2818063},
journal = {ACM Trans. Graph.},
month = {nov},
articleno = {246},
numpages = {9}
}

@article{weiss2019crowds,
title = {Position-based real-time simulation of large crowds},
journal = {Computers \& Graphics},
volume = {78},
pages = {12-22},
year = {2019},
issn = {0097-8493},
doi = {https://doi.org/10.1016/j.cag.2018.10.008},
url = {https://www.sciencedirect.com/science/article/pii/S0097849318301699},
author = {Tomer Weiss and Alan Litteneker and Chenfanfu Jiang and Demetri Terzopoulos},
}

@article{xing2022surface,
author = {Xing, Jingrui and Ruan, Liangwang and Wang, Bin and Zhu, Bo and Chen, Baoquan},
title = {Position-Based Surface Tension Flow},
year = {2022},
issue_date = {December 2022},
publisher = {Association for Computing Machinery},
address = {New York, NY, USA},
volume = {41},
number = {6},
issn = {0730-0301},
url = {https://doi.org/10.1145/3550454.3555476},
doi = {10.1145/3550454.3555476},
journal = {ACM Trans. Graph.},
month = {nov},
articleno = {244},
numpages = {12}
}

@inproceedings{yu2024xpbi,
author = {Yu, Chang and Li, Xuan and Lan, Lei and Yang, Yin and Jiang, Chenfanfu},
title = {XPBI: Position-Based Dynamics with Smoothing Kernels Handles Continuum Inelasticity},
year = {2024},
isbn = {9798400711312},
publisher = {Association for Computing Machinery},
address = {New York, NY, USA},
url = {https://doi.org/10.1145/3680528.3687577},
doi = {10.1145/3680528.3687577},
booktitle = {SIGGRAPH Asia 2024 Conference Papers},
articleno = {65},
numpages = {12},
location = {
},
series = {SA '24}
}

@article{10.1145/3197517.3201395,
author = {Chen, Hsiao-Yu and Sastry, Arnav and van Rees, Wim M. and Vouga, Etienne},
title = {Physical simulation of environmentally induced thin shell deformation},
year = {2018},
issue_date = {August 2018},
publisher = {Association for Computing Machinery},
address = {New York, NY, USA},
volume = {37},
number = {4},
issn = {0730-0301},
url = {https://doi.org/10.1145/3197517.3201395},
doi = {10.1145/3197517.3201395},
journal = {ACM Trans. Graph.},
month = {jul},
articleno = {146},
numpages = {13}
}

@inproceedings{SMPL-X:2019,
  title = {Expressive Body Capture: {3D} Hands, Face, and Body from a Single Image},
  author = {Pavlakos, Georgios and Choutas, Vasileios and Ghorbani, Nima and Bolkart, Timo and Osman, Ahmed A. A. and Tzionas, Dimitrios and Black, Michael J.},
  booktitle = {Proceedings IEEE Conf. on Computer Vision and Pattern Recognition (CVPR)},
  pages     = {10975--10985},
  year = {2019}
}

@inproceedings{bertiche2020cloth3d,
                    title={CLOTH3D: clothed 3d humans},
                    author={Bertiche, Hugo and Madadi, Meysam and Escalera, Sergio},
                    booktitle={European Conference on Computer Vision},
                    pages={344--359},
                    year={2020},
                    organization={Springer}
    }

@article{10.1145/3731183,
author = {Lan, Lei and Lu, Zixuan and Yuan, Chun and Xu, Weiwei and Su, Hao and Wang, Huamin and Jiang, Chenfanfu and Yang, Yin},
title = {JGS2: Near Second-order Converging Jacobi/Gauss-Seidel for GPU Elastodynamics},
year = {2025},
issue_date = {August 2025},
publisher = {Association for Computing Machinery},
address = {New York, NY, USA},
volume = {44},
number = {4},
issn = {0730-0301},
url = {https://doi.org/10.1145/3731183},
doi = {10.1145/3731183},
journal = {ACM Trans. Graph.},
month = jul,
articleno = {44},
numpages = {15}
}

\appendix
\section{Equivalence of XPBD and GPBD updates for linearly compliant constraints}
\label{app:equivalence}

Since XPBD and GPBD consider only one constraint at a time, we drop the subscript $i$ in this section.

Consider a constraint function $c(\bfx)$ with compliance $\alpha$ and associated energy $U(\bfx) = \frac1{2\alpha}c(\bfx)^2$.
The XPBD algorithm maintains a Lagrange multiplier $\lambda$ which is used to update the positions $\bfx$ on each iteration according to
\begin{align}
  \Delta\lambda &= -\frac{c(\bfx)+\Dt^{-2}\alpha\lambda}{\nabla c^T\bfM^{-1}\nabla c+\Dt^{-2}\alpha}, \\
  \lambda &\gets \lambda + \Delta\lambda, \\
  \bfx &\gets \bfx + \bfM^{-1}\Delta\lambda\nabla c(\bfx).
\end{align}

On the other hand, in GPBD we track the displacement $\bfd$ caused by the constraint, and update it according to
\begin{align}
  \bfd &\gets \bfd + \Delta\bfd, \\
  \bfx &\gets \bfx + \Delta\bfd, \\
  \text{s.t. }\bfd+\Delta\bfd &= -\bfW\nabla U(\bfx+\Delta\bfd).
\end{align}
For the case of linearly compliant constraints, the last equation becomes
\begin{align}
\bfd+\Delta\bfd &= -\alpha^{-1}\bfW c(\bfx+\Delta\bfd)\nabla c(\bfx+\Delta\bfd).
\end{align}
Adopting the XPBD assumption that the constraint Hessian can be ignored, we can derive the Newton update
\begin{align}
\bfd+\Delta\bfd &= -\alpha^{-1}\bfW (c(\bfx) + \nabla c^T\Delta\bfd)\nabla c \\
\implies (\bfI + \alpha^{-1}\bfW\nabla c\nabla c^T)\Delta\bfd &= -(\alpha^{-1}\bfW c(\bfx)\nabla c + \bfd).
\end{align}

Suppose the current GPBD guess for $\bfd$ is consistent with the constraint force direction, i.e. there is a Lagrange multiplier $\lambda$ such that $\bfd = \bfM^{-1}\lambda\nabla c(\bfx) = \Dt^{-2}\bfW\lambda\nabla c(\bfx)$.
Then, one may observe that the GPBD update $\Delta\bfd$ satisfies
\begin{align}
\Delta\bfd + \alpha^{-1}\bfW(\nabla c^T\Delta\bfd)\nabla c = -\alpha^{-1}\bfW c(\bfx)\nabla c - \bfd.
\end{align}
Here all the terms except the first are scalar multiples of $\bfW\nabla c$, so $\Delta\bfd$ must be as well.
Hence, we can write $\Delta\bfd = \Dt^{-2}\bfW\Delta\lambda\nabla c$ for some $\Delta\lambda$.
At this point, comparing coefficients of $\bfW\nabla c$, we are left with the scalar equation
\begin{align}
\Dt^{-2}\Delta\lambda + \Dt^{-2}\alpha^{-1}(\nabla c^T\bfW\nabla c)\Delta\lambda = -\alpha^{-1}c(\bfx) - \Dt^{-2}\lambda,
\end{align}
which can be easily simplified to the XBPD update rule.


\end{document}